\documentclass[%
 amsmath,amssymb,
 aps,
rmp,
showkeys,
]{revtex4-2}

\usepackage{graphicx}
\usepackage{dcolumn}
\usepackage{bm}
\usepackage[hidelinks]{hyperref}

\setcitestyle{super,comma,sort&compress}
\begin{document}
\title{Measles transmission and surveillance in open subcommunities}

\author{Niket Thakkar}
\email{niket.thakkar@gatesfoundation.org}
\homepage[\\]{https://nthakkar.github.io/}
\affiliation{%
 The Gates Foundation's Institute for Disease Modeling\\
 Seattle, Washington
}%
\date{\today}

\begin{abstract}
We're often in a situation where we have models of measles transmission that are stuck at a coarse population scale. This paper is about increasing their resolution or, said differently, creating consistent models for the population's subcommunities. We pay special attention to the fact that even if the full population is isolated, subcommunities are open to measles importations and related consequences of population movement.

\quad It turns out that the resolution issue can be productively framed as a constrained optimization problem, where the constraint is aggregation to the coarse model. That mathematical scaffolding allows us to make epidemiologically motivated approximations concentrated at the large population scale, where they're more valid, which leaves the estimates at the fine subcommunity scale atypically unstructured.

\quad We develop the idea in the context of the pre-vaccination-era 60 city UK dataset, building transmission models for all 60 cities. We find that they have intuitive seasonality, volatility, and importation elements, but that they depend on nontrivial dynamics in their case reporting probabilities. Potential interpretations of those surveillance processes are discussed in some detail, and we also describe how they influence inference on the balance between competing cryptic transmission and transient elimination dynamics. 
\end{abstract}
\maketitle

\section{\label{sec:intro}Introduction}
Measles immunization programs have understandable ambition to target vaccine delivery with increasing precision, vaccinating children on time more often and with smaller-scale interventions\cite{mri_strat,lindstrand2024implementing}. Recent financial pressures on public health have only added new urgency to these goals. From the quantitative science perspective, supporting targeted delivery means we need to be able to characterize epidemiologies at finer population scales, where we expect dynamics to be more volatile and data to be more sparse. 

This has generally been a challenging problem. Measles transmission models often operate at scales large enough where researchers feel safe assuming transmission is endemic in an isolated population \cite{finkenstadt2000time,ferrari2008dynamics,verguet2015controlling,thakkar2019decreasing}. Under that assumption, periods with zero measles cases are always interpreted as periods of cryptic transmission as opposed to periods of genuine interruption and elimination, making data sparsity more tractable. That zeros get reported at all illustrates that transmission is only partially observed, and researchers work towards a logical balance between variation in transmission and surveillance consistent with the data they have.

In smaller communities, transient elimination and reintroduction adds a new dynamic, one that forces us to think about connections between populations. When it comes to reported zeros we have some intuition. For example, local transmission is probably more common than imported transmission, multiple reported zeros in a row is a stronger indication of elimination than isolated zeros, transmission rates vary seasonally with common factors across scales, and so on. Building finer transmission models means making this intuition quantitative.

In the literature, this motivates importation models that range in complexity, from deterministic constants\cite{zimmerman2019optimization} to stochastic, economics-inspired models of population movement\cite{capoani2023review,bjornstad2019comparison}. As these models become more intricate, specifying them increases data demands, and that's led researchers to leverage various correlates of community connections, like air-traffic\cite{truelove2020epidemics} or cell-phone location data\cite{kostandova2025comparing,kostandova2025adjusting}. It's not a particularly satisfying solution -- that data is inconsistently available, often proprietary if it exists, and only tenuously connected to childhood diseases like measles.

Regardless of how we feel about cell-phone tracking kids, one conspicuous feature of the current state of the art is that endemic transmission models, those that operate at the large-population scale, are essentially disconnected from models at finer resolution. In practice, we decide first what scale to operate at, choose  mechanisms accordingly, and never check directly on aggregation properties and the recovery of a specific endemic limit. 

Inspired by this context, this paper describes an alternative approach to a simplified, related problem. Specifically, we consider a situation where we have an endemic transmission model we believe is accurate, and we conditionally infer a likely disaggregation of that model consistent with data from a single subcommunity of interest. We discuss the details by way of example using the canonical, pre-vaccination-era, 60 city UK dataset\cite{finkenstadt2000time,bjornstad2002dynamics,grenfell2002dynamics}, with any single city as the subcommunity. At a high level, by characterizing constrained differences between the 60 city model and a model of the remaining 59, we can infer consistent transmission, importation, and surveillance dynamics that fit the data from all cities and scale intuitively with their size.

As we'll see, solving the subcommunity problem in this way exposes the potentially underappreciated role of surveillance probability changes in warping time series at different population scales. We discuss this finding in some detail, showing that the inferred reporting probabilities are consistent with an excitable, nonlinear system reacting to stimulus from transmission, motivating the need for more nuanced understanding of measles surveillance systems generally. We also address the issue of interpreting time periods of zero case reports with these inferences in hand, and we find that asymmetries in the surveillance dynamics have an outsized effect on interruption probabilities. 

Taken as a whole, this paper covers three main ideas: (1) That aggregation to a known endemic model tractably constrains subcommunity epidemiologies, (2) that reporting probabilities with significant, interpretable variation arise naturally in the absence of restrictive models, and (3) that zeros in cases can be interpreted contextually when transmission, importation, and surveillance processes are specified. That this can all be accomplished without direct mechanistic reference to population movement is interesting as well, and we comment on that throughout.

\section{\label{sec:table}An inference problem}
To get started let's state the inference problem more concretely. In Ref. \citenum{thakkar2022modeling}, we developed a discrete time, stochastic process model of measles transmission across the 60 UK cities from $1944$ to $1966$, before measles vaccine was introduced in $1967$. That model compartmentalized the population into infectious, $I_t$, and susceptible, $S_t$, states, interacting as
\begin{align}
    I_{t+1} &= \tilde{\beta}_t S_t I_t^{\alpha}, \label{eq:I} \\
    S_{t+1} &= S_{t} + B_t - I_{t+1},\label{eq:S}\\
    C_t &\sim \text{Binom}\left\{I_t,\,r_t\right\},\label{eq:C}
\end{align}
with a seasonal, log-normally distributed transmission rate $\tilde{\beta}_t$, a saturation parameter $\alpha \leq 1$, and a binomial surveillance process with reporting probability $r_t$, all evolving on 2-week time steps, $t$, approximating the measles generation time. The model was fit to reported cases, $C_t$, leveraging recorded births, $B_t$, and a published age-distribution of measles cases from 1950. 

\begin{figure*}
\centering\includegraphics[width=\linewidth]{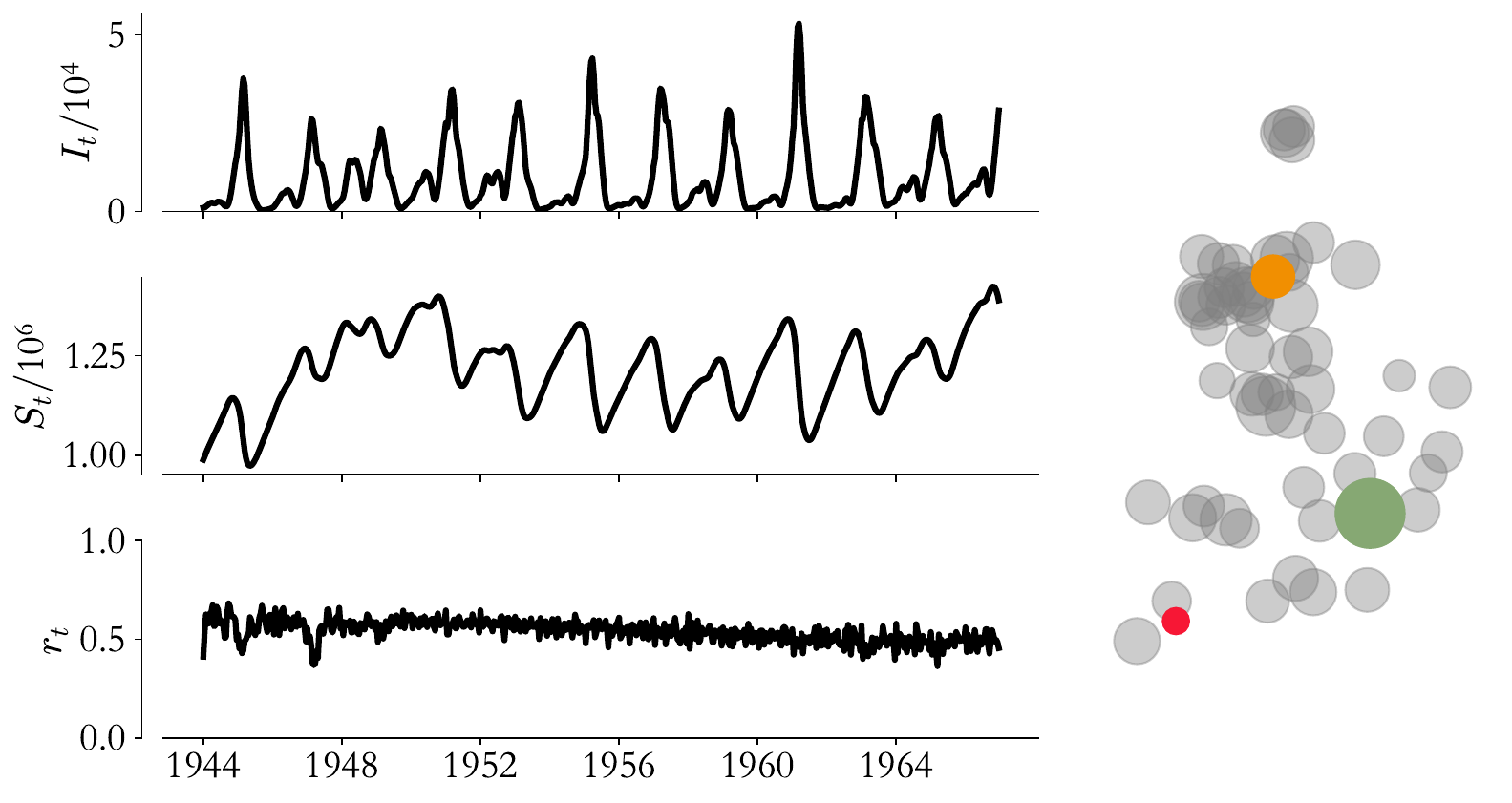}
\caption{\label{fig:int_pt} Measles in the pre-vaccine-era UK. (right) The data comes from 60 cities located across England and Wales (grey dots). Throughout the paper, we use the largest (London, green), smallest (Teignmouth, red), and median (Huddersfield, orange) cities as representative examples. (left) Expected $I_t$, $S_t$, and $r_t$ for all 60 cities in aggregate was estimated previously\cite{thakkar2022modeling}. The $r_t$ series is roughly constant if not for the dips that vanish after 1948, an effect we hypothesized corresponds to the transition to universal healthcare at that time.}
\end{figure*}

Eqs. \ref{eq:I} to \ref{eq:C} are closely related to classic models in the literature\cite{finkenstadt2000time} and analyzed in detail in Ref. \citenum{thakkar2022modeling}. We take them as a starting point here, and in Fig. \ref{fig:int_pt}, we visualize the model's expected states for reference. Note that $I_t$ can never be zero by the log-normal assumption, and the population is isolated, so this an endemic model in the language of the previous section.

Now we imagine learning that the data is disaggregated into 2 pieces, $C_t = C_{+t} + C_{-t}$ for example, where $C_{+t}$ are cases from a subcommunity of specific interest and $C_{-t}$ is the remainder, what we'll call the $+$community (the ``plus community") and $-$community (the ``minus community") from now on. In the UK example we're working through, the $+$community is always a single city and the $-$community is always the remaining 59, but other, even non-spatial, disaggregations are possible in principle. The project we take up is estimating $I_{+t}$, $S_{+t}$, and $r_{+t}$ conditional on the time series in Fig. \ref{fig:int_pt} in light of this new information.

\section{\label{sec:feas}The endemic model as a constraint}
It's useful to specify how the $+$ and $-$communities are related in the context of a fixed endemic model. We know $I_t = I_{+t} + I_{-t}$ and $S_t = S_{+t} + S_{-t}$ $\forall t$ if the full population is isolated and subcommunity membership is assigned at birth. Meanwhile, taking $C_{\pm t} \sim \text{Binom}\left\{I_{\pm t}, r_{\pm t}\right\}$, implies via $\text{E}[C_{\pm t}|I_{\pm t}, r_{\pm t}] = r_{\pm t}I_{\pm t}$ and Eq. \ref{eq:C} that
\begin{align*}
    r_t = r_{+t}\frac{I_{+t}}{I_t} + r_{-t}\frac{I_{-t}}{I_t},
\end{align*}
an intuitive, prevalence weighted average. Unsurprisingly, we can estimate the $+$community states by focusing on the $-$community states, which we might better understand since they represent a larger population. 

That said, we do have to resist the temptation to fit a version of Eqs. \ref{eq:I} to \ref{eq:C} to the $-$community data carelessly. To see why, note that $I_{+t} \geq C_{+t}$ on the discrete $t$ timescale. Rearranging that inequality, making use of $\text{E}[I_{-t}|r_{-t},C_{-t}] = (C_{-t}+1-r_{-t})/r_{-t}$ for binomial $C_{-t}$ (see Appendix B in Ref. \citenum{thakkar2022modeling}), gives
\begin{align}
    r_{-t} \geq \frac{C_{-t}+1}{\text{E}[{I}_t|r_{t}, C_t]+1 - C_{+t}} \equiv \ell_{-t},\label{eq:lb}
\end{align}
in expectation, where $\text{E}[{I}_t|r_{t}, C_t]$ is based on the endemic model, Fig. \ref{fig:int_pt}'s top panel for the UK example. Eq. \ref{eq:lb} tells us that valid models of the $-$community dynamics have $r_{-t}$ bounded below because of feasibility constraints on the $+$community states. Said differently, if $r_{-t}$ is too low, $C_{-t}$ corresponds to too many $-$community infections, and since $I_t = I_{+t} + I_{-t}$ is fixed, $I_{+t}$ becomes too small to explain $+$community cases. We'll have to make inferences mindful of these types of interactions.

Inspired by this observation, it's reasonable to search for more constraints, particularly an expected upper bound on $r_{-t}$. It turns out we get a useful one considering new infections, $N_t$, in the endemic model,
\begin{align*}
    N_t &\equiv \tilde{\beta}_t S_t I_t^{\alpha},\\
    &= \frac{\tilde{\beta}_t}{I_t^{1-\alpha}} S_t I_t \equiv \beta_t S_t I_t,
\end{align*}
where we've defined $\beta_t$ implicitly. In a closed population, $N_t$ is made up of 4 types of events,
\begin{align*}
    \beta_t S_t I_t =& \,\,\beta_{++t}S_{+t}I_{+t} + \beta_{--t}S_{-t}I_{-t}\\
        &+ \beta_{+-t}S_{+t}I_{-t} + \beta_{-+t}S_{-t}I_{+t},
\end{align*}
where $\beta_{ijt}$ is the realized fraction of all possible $S_{it}I_{jt}$ pairs for $i,j = +,-$. The contributions in the top line correspond to local transmission in the $+$ and $-$communities respectively while the second line's terms represent importation and exportation from the $+$community perspective. Rearranging gives
\begin{align}
    \beta_t S_t I_t - \beta_{--t}S_{-t}I_{-t}\left[1 + \frac{\beta_{-+t}I_{+t}}{\beta_{--t} I_{-t}}\right] &= \beta_{++t}S_{+t}I_{+t} + \beta_{+-t}S_{+t}I_{-t} \geq 0, \label{eq:gen_ub}
\end{align}
which exposes that transmission relevant to the $+$community is the difference between overall transmission and events relevant to the $-$community, a difference that cannot be negative. 

Eq. \ref{eq:gen_ub} is general but not particularly approachable. If we focus on situations where the $+$community is much smaller than the $-$community, we can assume that exportations contribute negligibly to $-$community transmission and the term in brackets is $\approx1$. Moreover, along similar lines, we assume the $+$community is small enough that seasonality is preserved in the $-$community: $R_e = \langle S_t \rangle\beta_{t} \approx \langle S_{-t} \rangle\beta_{--t}$, where $\langle\cdot\rangle$ is the time average and $R_e$ is called the effective reproductive number. We expect $\langle S_{-t}\rangle/\langle S_{t}\rangle \approx F_{-}$, the average fraction of the population within the $-$community, since endemic susceptibility levels vary slowly across population scales\cite{anderson1992infectious,bjornstad2002dynamics}. In this context, Eq. \ref{eq:gen_ub} simplifies,
\begin{align*}
    F_{-} S_t I_t - S_{-t}I_{-t} \geq 0,
\end{align*}
and taking $S_{-t+1} = S_{-t} + B_{-t} - I_{-t+1}$ while again making use of $\text{E}[I_{-t}|r_{-t},C_{-t}]$ shows that the conditional expectation of left-hand-side is specified given the data, initial condition $S_{-0}$, and a candidate reporting probability series. That means, with boldface indicating vector time series, 
\begin{align}
h_t(\mathbf{r}_{-t},S_{-0}) = F_{-} \text{E}[{S}_t{I}_t|\mathbf{r}_t,\mathbf{C}_t,\mathbf{B}_t] - \text{E}[S_{-t}I_{-t}|\mathbf{r}_{-t},\mathbf{C}_{-t},\mathbf{B}_{-t}]\geq 0,\label{eq:ub}
\end{align}
is a nonlinear and nonlocal constraint on $r_{-t}$ at fixed $S_{-0}$. 

We prove algebraically that $h_t(\mathbf{r}_{-t},S_{-0}) \geq 0$ is in fact an upper bound on $r_{-t}$ in Appendix \ref{app:A}, but more intuitively, if $r_{-t'} = u_{-t'}$, a potential upper bound for $t' < t$, then $S_{-t}$ is as large as possible since past cases correspond to as few infections as possible. Then, if $r_{-t}$ suddenly falls to $\ell_{-t}$, $I_{-t}$ is as large as possible, which maximizes $S_{-t}I_{-t}$ and minimizes $h_t(\mathbf{r}_{-t},S_{-0})$. Thus maintaining $h_t(\mathbf{r}_{-t},S_{-0}) \geq 0$ bounds susceptibility accumulation from above which sets $u_{-t'}$. Epidemiologically, if reporting in the $-$community is too high, $S_{-t}$ accumulates and outbreaks have the potential to be more explosive than is allowed by the endemic model. 

Both bounds $\ref{eq:lb}$ and \ref{eq:ub} can be evaluated at any candidate $\mathbf{r}_{-t}$, since the expected $S_{-t}$, $I_{-t}$, and $S_{-t}I_{-t}$ can be constructed directly from $\mathbf{C}_{-t}$ and $\mathbf{B}_{-t}$ when reporting rates and initial conditions are fixed (see Appendix \ref{app:A}). As Eq. \ref{eq:lb} is a lower bound and Eq. \ref{eq:ub} is an upper bound, they define a region of feasible $r_{-t}$ values. We're left to find the best choice in that region.

In Appendix \ref{app:B}, we show that 
\begin{align}
    I_{-t+1} &= \frac{\beta_t}{F_{-}}S_{-t}I_{-t},\label{eq:trans_minus}\\
    I_{+t+1} &= \beta_t S_t I_t - \frac{\beta_t}{F_{-}}S_{-t}I_{-t},\label{eq:trans_plus}
\end{align}
define an autoregressive posterior probability distribution $\text{pr}(\mathbf{r}_{-t}\,|\,S_{-0},\mathbf{C}_{\pm t},\mathbf{B}_{\pm t})$, where conditionality on the endemic model is implicit. Note specifically that transmission in the $+$community, Eq. \ref{eq:trans_plus}, is based on what's left over from the $-$community, allowing us to avoid direct assumptions about $\beta_{++t}$ and $\beta_{+-t}$. Our project is now much more specific,
\begin{align*}\tag{$\mathcal{P}$}\label{eq:nlp}
    \begin{split}
    \max_{\mathbf{r}_{-t}}& \quad \text{pr}(\mathbf{r}_{-t}\,|\,S_{-0},\mathbf{C}_{\pm t},\mathbf{B}_{\pm t})\\
    \text{subject to}&\quad r_{-t} \geq \ell_{-t}\\
    &\quad h_t(\mathbf{r}_{-t},S_{-0}) \geq 0\,\,\forall t,
    \end{split}
\end{align*}
a so-called constrained, nonlinear program for $\mathbf{r}_{-t}$ and downstream conditional expectations for $I_{\pm t}$, $S_{\pm t}$, and $r_{+t}$. \ref{eq:nlp} encapsulates the idea that we're looking for expected states that best explain observed cases while maintaining a seasonal transmission process and aggregating correctly to the endemic model. 

In Appendix \ref{app:A}, we develop an iterative, interior point algorithm for programs like \ref{eq:nlp}. At a high level, we start at $r_{-t}^{(0)} = \ell_{-t}$, approximate $u_{-t}^{(0)}$ by linearizing Eq. \ref{eq:ub} around $r_{-t}^{(0)}$, and then maximize $\text{pr}(\mathbf{r}_{-t}\,|\,S_{-0},\mathbf{C}_{\pm t},\mathbf{B}_{\pm t})$ in the region $\ell_{-t} \leq r_{-t} \leq u_{-t}^{(0)}$. That result defines $r_{-t}^{(1)}$, and then this process is repeated until the maximum probability value changes less than some preset tolerance. In practice, on a laptop, the full process takes less than a couple minutes.

\begin{figure*}
\centering\includegraphics[width=\linewidth]{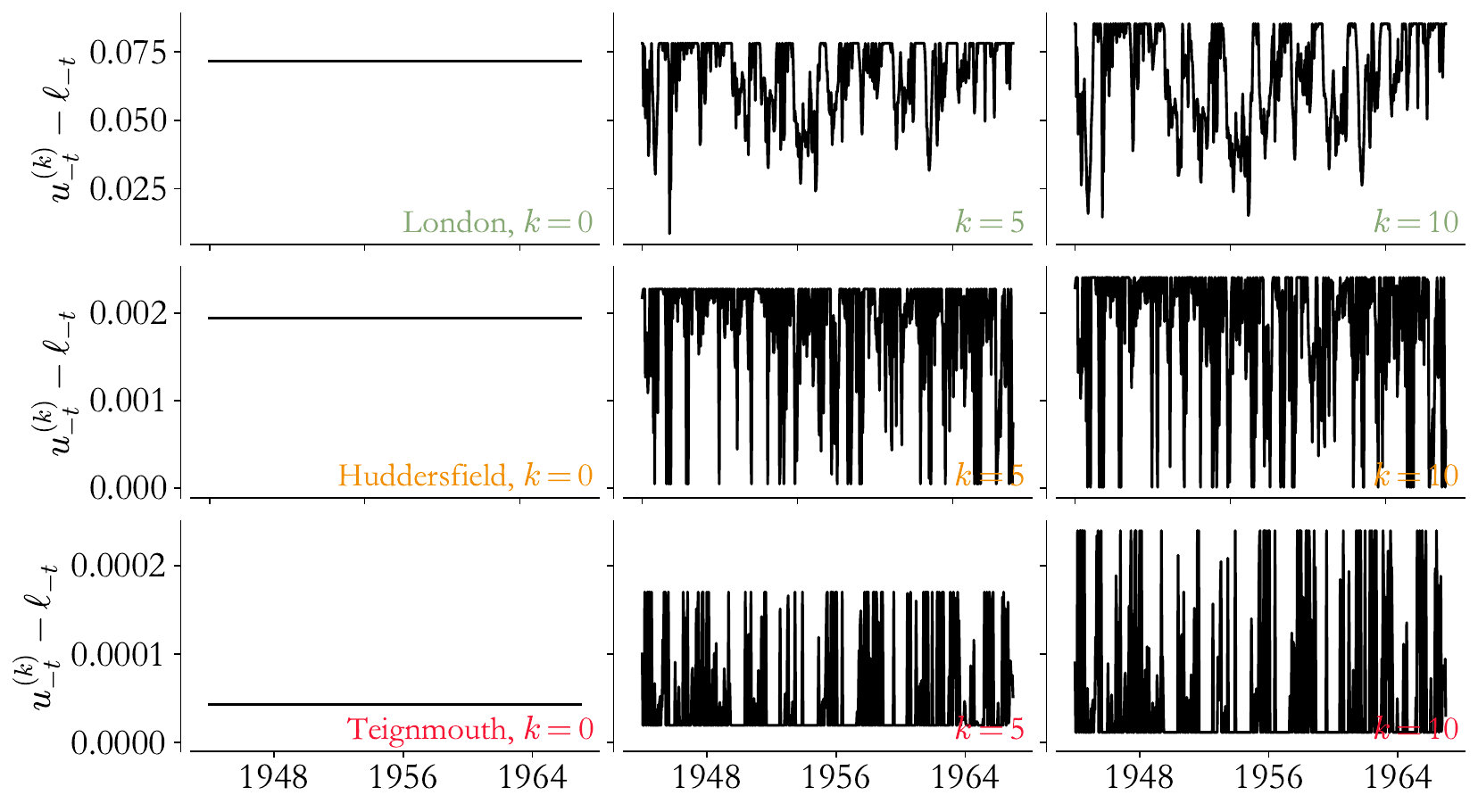}
\caption{\label{fig:feas} Feasible changes in $r_{-t}$. When we iteratively estimate how much $r_t$ can change if a specific city is removed, we show that feasibility is a complex function of the data, but also that a city's influence on the whole decreases quickly with its size.}
\end{figure*}

We can see some interesting properties of the optimization process by visualizing the approximate feasible region, $u_{-t}^{(k)}-\ell_{-t}$, for a few iterations, $k$. This is done in Fig. \ref{fig:feas} for 3 cities, London, Huddersfield, and Teignmouth, the largest, median, and smallest of the 60 respectively. Through Eqs. \ref{eq:lb} and \ref{eq:ub}, the regions are informed by observed cases and births, and they inherit complex structure from that data. We also see that as iteration progresses, the maximum allowable change in $r_{-t}$ grows. This is because Eq. \ref{eq:ub} is nonlocal, meaning that increased restriction in one time period relaxes constraints in another. Epidemiologically, $S_{-t}$ keeps long term memory from past outbreaks, so if a past outbreak's size decreases, present day susceptibility accommodates more outbreak size variation.

A key feature of Fig. \ref{fig:feas} is that as $+$community size decreases, the overall feasible region gets smaller. For example, London contains about $20\%$ of the 60 city population, and $r_{-t}$ can potentially change from the overall $r_{t}$ by roughly $5\%$ as we modify estimates in London. Meanwhile, Huddersfield contains $<1\%$ of the full population, and $r_{-t}$ can shift from $r_t$ by only about $0.1\%$. In other words, as the $+$community gets smaller, \ref{eq:nlp} is more restricted, with better underlying assumptions, and so is easier to solve. If we had started from a mobility model of community connection, this is a little surprising, as those models become more volatile and less well-defined at smaller population scales. 

\begin{figure*}
\centering\includegraphics[width=\linewidth]{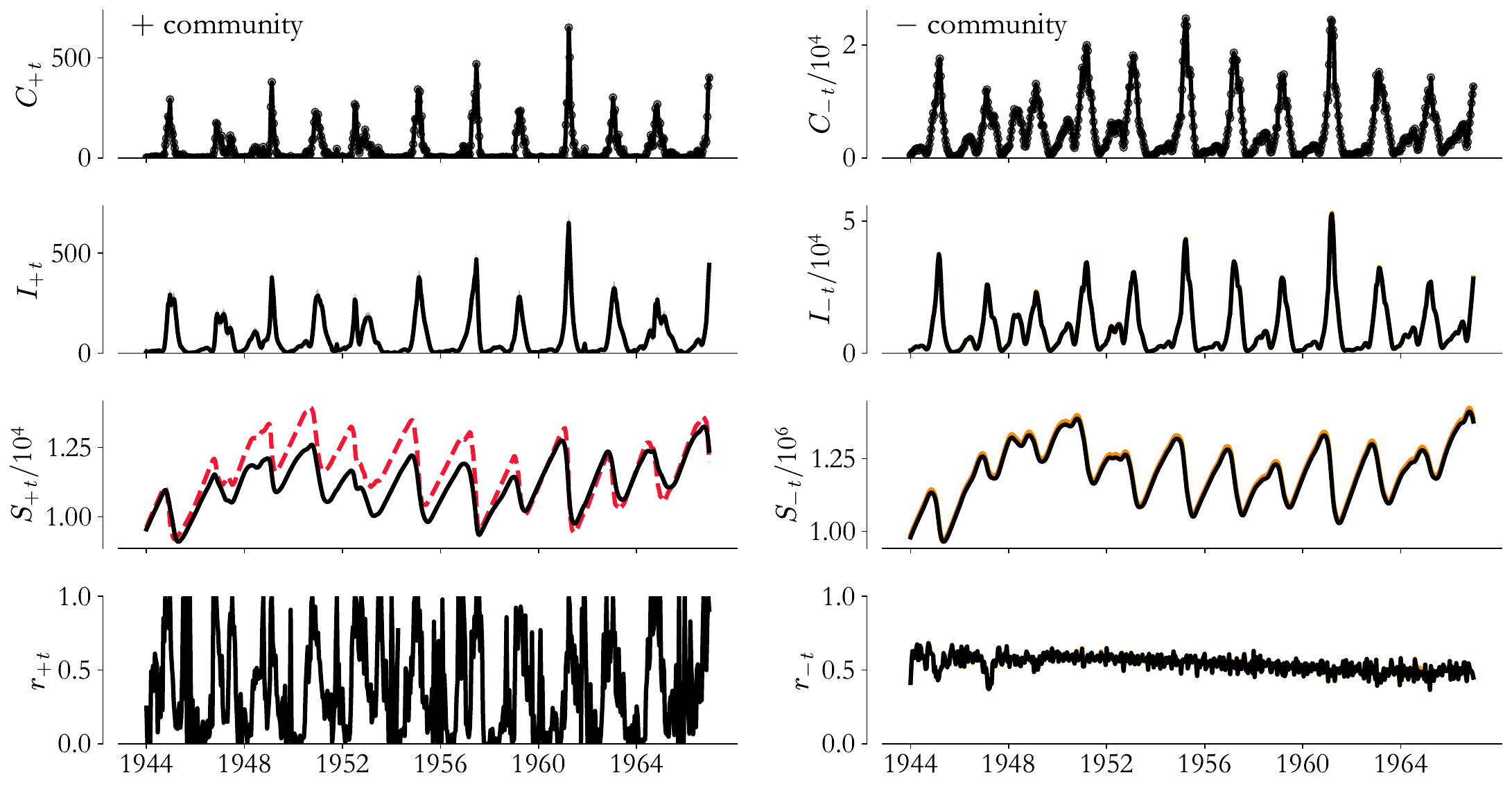}
\caption{\label{fig:estimates} Huddersfield as an example. Estimates for the 59 cities excluding Huddersfield (black, right) barely change from those for the full collection (orange). The difference specifies the states in Huddersfield (black, left). While the inferences fit cases (black circles) well, they do so with a highly dynamic reporting probability in Huddersfield.}
\end{figure*}

In any case, we can solve \ref{eq:nlp}. An example solution is visualized for Huddersfield in Fig. \ref{fig:estimates}, comparing the $+$ and $-$community dynamics to the total and each other. The inferred $-$community states have nearly imperceptible deviations from the endemic model (orange). We compute those differences, that is the $+$community states, and visualize them on the left. The estimates, essentially highly structured residuals of \ref{eq:nlp}, fit $C_{+t}$ with an underlying $S_{+t}$ that qualitatively agrees with more conventional estimates from the literature\cite{bjornstad2002dynamics,grenfell2002dynamics} (red dashed line). That said, these expected estimates aggregate to those in Fig. \ref{fig:int_pt} exactly.

A main distinction between these estimates and others is the inferred structure and volatility in $r_{+t}$. We find that reporting probabilities in Huddersfield vary from 0 to 100$\%$, with visually apparent time correlation. Across the 60 cities, this finding holds qualitatively, and it speaks to the lack of assumptions we're making regarding possible dynamics in the $+$community. But we shouldn't accept this estimate unless we can plausibly explain it. That's the basic project of the next two sections.

\section{\label{sec:trans}Subcommunity transmission}


Results like those in Fig. \ref{fig:estimates} give us non-mechanistic perspective on $+$community transmission, in the sense that Eq. \ref{eq:trans_plus} avoids reference to individuals in $S_{+t}$ transitioning to infectiousness. While we can compute signals like $\text{E}[\beta_tS_t I_t - (\beta_t/F_{-})S_{-t}I_{-t}]$ at this stage, which tells us the expected number of transmission events in the $+$community, we have yet to uncover any corresponding relationship between $I_{+t}$ and $S_{+t}$.

Depending on our goals, we could test a variety of mechanistic models against the point estimates we get by solving \ref{eq:nlp}, but considering the $+$community contributions in Eq. \ref{eq:gen_ub}, an approachable starting point is
\begin{align}
    I_{+t+1} = N_{+t} \sim \text{NegBin}\left\{\text{E}\left[\gamma\beta_{t}S_{+t}\left(I_{+t} + \lambda I_{-t}\right)\right],\eta\right\},\label{eq:trans_mod}
\end{align}
where $\eta$ accommodates over-dispersion relative to a Poisson distribution. Here, we've assumed $\beta_{++t} \approx \gamma \beta_{t}$ and $\beta_{+-t}/\beta_{++t}\approx\lambda$, that the local transmission rate is directly proportional to that from the endemic model and that importations are directly proportional to infections in the $-$community. These are simplistic models of the transmission components, and in the UK example they turn out to perform well. But it's worth emphasis that we could test any model of interest against the point estimates we get by solving \ref{eq:nlp}.

To specify the proportionality constants in Eq. \ref{eq:trans_mod}, we compute the non-negative projection of
the expected number of transmission events above against $\text{E}[\gamma\beta_{t}S_{+t}\left(I_{+t} + \lambda I_{-t}\right)]$ with standard methods\cite{lawson1995solving}. In the Huddersfield example, we find that the projection explains $99\%$ of the variation in $\text{E}[\beta_tS_t I_t - (\beta_t/F_{-})S_{-t}I_{-t}]$ with a locally dominated transmission process ($\langle\lambda I_{-t}\rangle/\langle I_{+t}\rangle \approx 5\%$), a result reproduced across the cities, strongly supporting our coarse choices for $\beta_{++t}$ and $\beta_{+-t}$. Moreover, given $\gamma$ and $\lambda$, we can find the $\eta$ that maximizes the autoregressive likelihood of Eq. \ref{eq:trans_mod}, fully specifying the mechanistic model. 

\begin{figure*}
\centering\includegraphics[width=\linewidth]{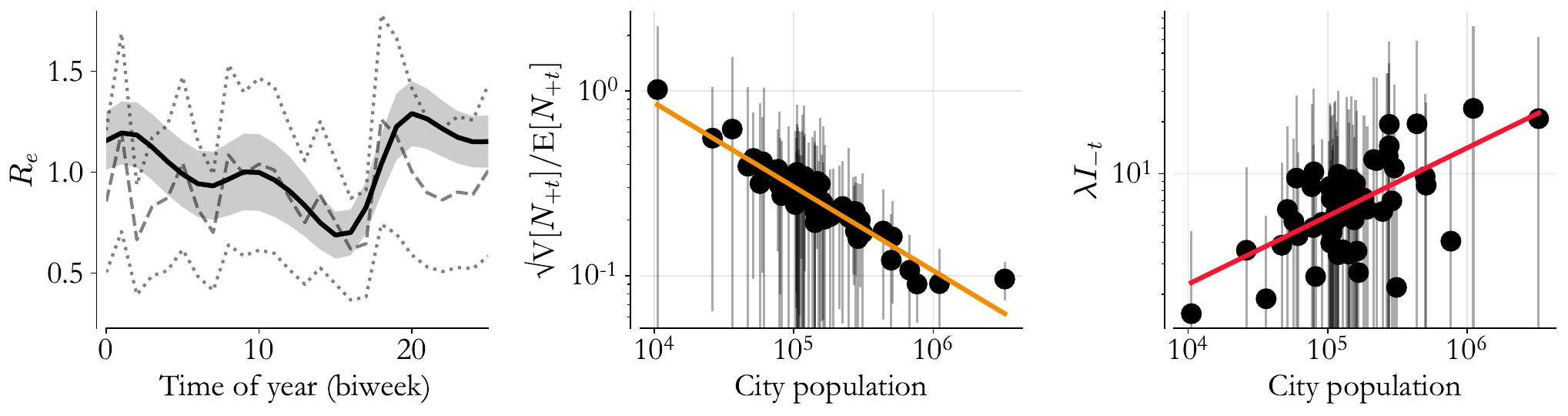}
\caption{\label{fig:trans}Measles transmission. (left) The estimates in Fig. \ref{fig:estimates} yield an $R_e$ with meaningful seasonality (black, 2 standard deviations shaded), as opposed to a model without reporting probability variation (grey, 2 standard deviations dotted). (middle) Transmission volatility still scales with population size (yellow). (right) Importation volume is highly seasonal (2 standard deviation error bars around the time average), but scales on average with $(0.40 \pm 0.14)\log_{10}(\text{size})$ (red). }
\end{figure*}

Some properties of these models across the 60 cities are visualized in Fig. \ref{fig:trans}. In the first panel, since $\beta_{++t} \approx \gamma \beta_{t}$, Eq. \ref{eq:trans_mod} inherits the average seasonality from the endemic model (black). That said, illustrated here for Huddersfield, volatility around $R_e$ (grey region), while significant, is less than the time-variation in the seasonality profile. In other words, the transmission process is meaningfully seasonal instead of variance dominated. 

For comparison, if we estimate $R_e$ from $C_{+t}$ directly under a more conventional, constant reporting probability model (see Appendix \ref{app:C}), we get the dashed-line estimate. While the average correlates with that from Eq. \ref{eq:trans_mod}, transmission process variance is much larger (dotted lines), generally covering $R_e=1$, making seasonal variation insignificant. This indicates that, in solving \ref{eq:nlp}, we've cleaned $C_{+t}$ of reporting artifacts to recover a much more well-defined, seasonal transmission dynamic in the $+$community.

Looking across cities in the second panel, the coefficient of variation, $\sqrt{\text{V}[N_{+t}]}/E[N_{+t}]$, represented as a time-average with time-variation errorbars (black), shows that the transmission process becomes more deterministic as $+$community size increases (yellow line). In other words, volatility around the seasonality profile scales intuitively, with larger populations having more predictable transmission processes, in qualitative agreement with other studies\cite{bjornstad2002dynamics} of this dataset.

Finally, in the third panel, we visualize the expected number of importations, again as a time-average with time-variation errorbars (black). Average importation volume grows at an approximately square-root of population rate (red line), in good agreement with theoretical predictions\cite{bartlett1960critical} based on transmission's long-term stability. However, the large errorbars illustrate that seasonality in the importation process, via change in $I_{-t}$, is a key element of Eq. \ref{eq:trans_mod}. This has potential epidemiological implications: If we were modeling importation dynamics from a population movement perspective, times of high movement may be out of phase with the high-measles transmission season, depending on the underlying drivers of transmission. The results here suggest that the key driver of importations, at least in this example, is measles prevalence in the surrounding area. 

Zooming out, we've used Eq. \ref{eq:trans_mod} to test for epidemiologically plausible relationships between the $S_{+t}$ and $I_{+t}$ estimates we uncovered by solving \ref{eq:nlp}. We've found that those signals are consistent with a seasonal, locally-dominated, transmission process with intuitive volatility and importation volume scalings. Since \ref{eq:nlp} was constructed to avoid assumptions about $+$community transmission, relying only on assumptions about the $-$community, it's reassuring to see the population biology so clearly reflected in our signal processing. 

\section{\label{sec:rep}Subcommunity surveillance}

The transmission processes just discussed, while intuitive, rely heavily on unconventional reporting probability dynamics. To explain why $r_{+t}$ looks the way it does, we take two distinct approaches in this section. First, we offer a plausible, mechanistic interpretation of the $r_{+t}$ estimate for Huddersfield. Then, separately, we visualize how all 60 $r_{+t}$ estimates scale with prevalence, revealing interpretable epidemiological regimes. 

When it comes to explaining the $r_{+t}$ estimate from Fig. \ref{fig:estimates}, relative to the transmission components, there's unfortunately much less literature and biology to fall back on. Somewhat speculatively then, we might guess that reporting systems have some key features: Communities can rest in states of relative unawareness, they can be excited into states of hyper-awareness, and they are driven by inciting transmission events. The abrupt changes in the time series in Fig. \ref{fig:estimates} suggests those features, particularly when viewed in parallel to $I_{+t}$.

These types of dynamics are characteristic of stochastically forced, nonlinear, excitable systems, studied closely\cite{CEBRIANLACASA2024six} in mathematical neuroscience and cardiology. Inspired by that literature, we change units to $x_t = \Phi^{-1}(r_{+t})$, using the cumulative, standard-normal distribution $\Phi(\cdot)$ so that $r_{+t}\in[0,1]$ is one-to-one with $x_{t}\in\mathbb{R}$. Our hypothesis, based on the qualitative features just discussed, is that 
\begin{align}
    \begin{split}
    x_{t+1} &= x_t + ax_t + bx_t^3 + A_t,\\
    A_{t+1} &= A_t + c(\bar{A}-A_{t}) + d(T_{+t}-1),\label{eq:fhn}
    \end{split}
\end{align}
where $a$ and $b$ characterize the restoring force in $x_t$, and $A_t$ is a measure of attention that relaxes at rate $c$ to $\bar{A}$ in the absence of external forcing. The system is driven by transmission events, $T_{+t} \sim \text{NegBin}\{\mu_{+t},k_{+t}\}$, with mean and dispersion set by moment-matching
\begin{align}
    N_{+t} &= \sum_{i=1}^{I_{+t}+\lambda I_{-t}}T_{+t,i},\label{eq:bp}
\end{align}
to Eq. \ref{eq:trans_mod} up to second order\cite{thakkar2022covid}, where $T_{+t,i}$ is the $i$th infectious individual's contributions to $N_{+t}$ (see Appendix \ref{app:D}). Note that the seasonal time-average of $\mu_{+t}$ is $R_e$ from Fig. \ref{fig:trans}, and $T_{+t}$ is essentially a stochastic generalization of the effective reproductive number. 

Eq. \ref{eq:fhn} is based on the well-studied FitzHugh-Nagumo model\cite{CEBRIANLACASA2024six}, which is the minimal dynamical system capable of nonlinear excitation. The first line allows $x_t$ to rise and fall asymmetrically, and the second line allows $A_t$ to slowly process and linger on transmission events. We can specify the parameters in Eq. \ref{eq:fhn} by writing them as a second-order autoregression in $x_t$ and fitting them to $\Phi^{-1}(r_{+t})$ from Huddersfield in the least squares sense.

\begin{figure*}
\centering\includegraphics[width=\linewidth]{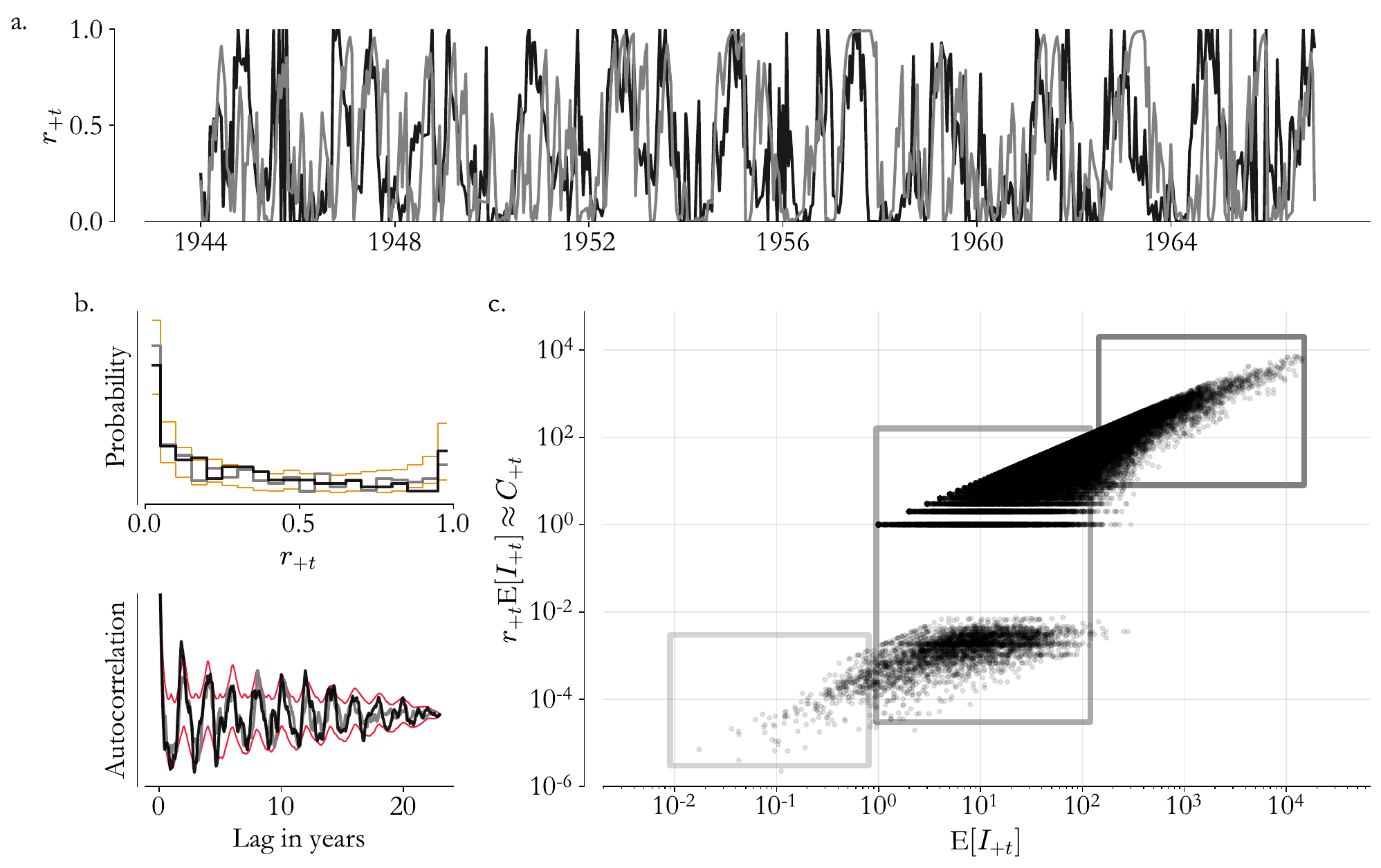}
\caption{\label{fig:rt}Measles surveillance. (a) The estimate from Fig. \ref{fig:estimates} (black) is similar to a sample from a transmission-driven, excitable system (grey) analogous to models of spiking neurons. (b) The excitable system (grey sample, 95$\%$ interval in color) explains the histogram and autocorrelation of the Huddersfield estimate (black). (c) $r_{+t}$ estimates across the 60 cities cluster into distinct prevalence regimes, suggesting that importation (light grey) and modest transmission chain detection (mid grey) are qualitatively different than reporting at higher burden (dark grey).}
\end{figure*}

The outcome of this analysis is visualized in Figs. \ref{fig:rt}a and b. In the top panel, the Huddersfield estimate (black) is compared to a full-time sample from the fitted model (grey). Both series are geometrically comparable across the 26 year period, with steep, asymmetric rises and falls at specific times of high transmission. In Fig. \ref{fig:rt}b, the excitable system is sampled thousands of times to estimate a histogram and autocorrelation function (95$\%$ interval in color). The estimate from Fig. \ref{fig:estimates} (black), generated by solving \ref{eq:nlp}, is statistically comparable to the visualized sample (grey). 

Fig. \ref{fig:rt} illustrates that our empirical estimate, based on regression in $r_{-t}$ without a specific relationship between $I_{+t}$, $S_{+t}$, and $r_{+t}$, is consistent with a mechanistic model driven by the transmission process in Eqs. \ref{eq:trans_mod} and \ref{eq:bp}. Specifically, we've found that a minimal dynamical system capable of transmission-driven excitation and relaxation creates time series that match $r_{+t}$ in distribution and in autocorrelation. This suggests that the reporting dynamics are not arbitrary artifacts of the optimization in \ref{eq:nlp}. They are in fact consistent with the biologically interpretable transmission process discussed in the previous section.

We could imagine studying Eq. \ref{eq:fhn} in much more depth, experimenting with certain mechanisms, analyzing properties of the parameters across cities, and so on. But since our main goal here is interpreting outputs of \ref{eq:nlp}, not commenting on surveillance in the UK specifically, we won't bother. Developing tractable models of modern surveillance systems does seem like it warrants further study, and FitzHugh-Nagumo-inspired dynamics are a potential starting point for that.

Moving on then, independent of Eq. \ref{eq:fhn}, Fig. \ref{fig:rt}c highlights that the $r_{+t}$ estimates across all 60 cities (black dots) cluster intuitively. Specifically, visualizing expected cases, $r_{+t}I_{+t}$, against $I_{+t}$ exposes distinct reporting regimes as the epidemiology changes during outbreaks. Emphasized by the grey boxes, from left to right, reporting probabilities are extremely low during importation detection periods, they become bimodal for moderate transmission chains, and then they limit to a well-defined average during periods of higher burden. 

These 3 phases give us insight into Fig. \ref{fig:estimates}'s stability in $r_{-t}$ and corresponding instability in $r_{+t}$. While the latter transitions between phases all the time, the former is large enough to be trapped at high prevalence, and the system has a well-defined average over the whole period as a result. This observed structure in $r_{+t}$ suggests that any transmission model outside the endemic limit should consider the effects of surveillance system regime switching.

\section{\label{sec:elim}Interruption probabilities}

The last two sections argued that estimates from \ref{eq:nlp} lead us to viable, consistent models for the transmission and surveillance processes in the $+$community. Recall that our motivating challenge was that cryptic transmission and transient elimination strike a complex balance outside the endemic limit, particularly when it comes to interpreting times when $C_{+t} = 0$. We describe that balance quantitatively in this section.

\begin{figure*}[b]
\centering\includegraphics[width=\linewidth]{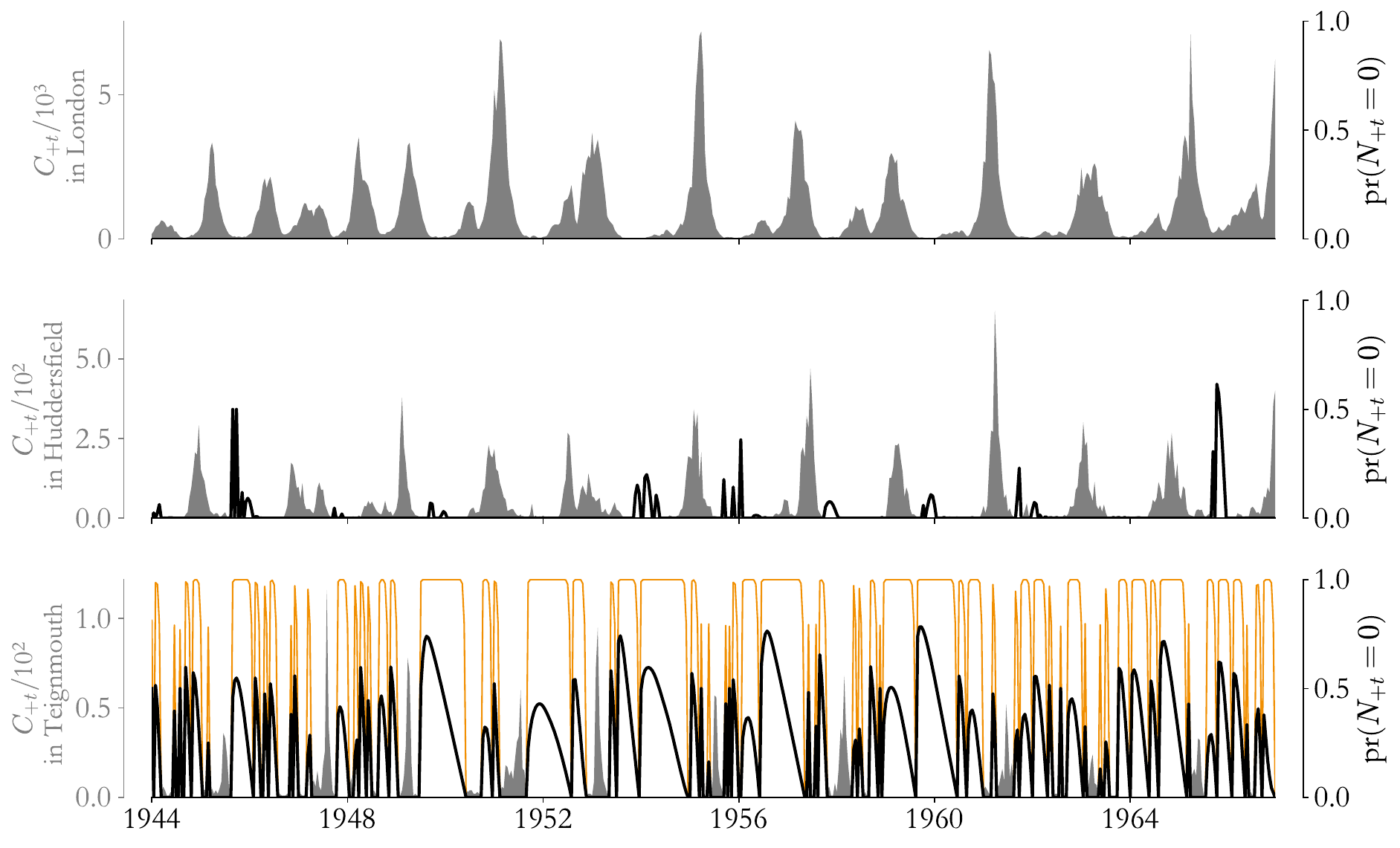}
\caption{\label{fig:elim}Interruption probabilities. The branching process perspective leads us to an efficient signal processing of cases (grey) into transmission interruption probabilities (black), shown here for London, Huddersfield, and Teignmouth. In Teignmouth, we compare to a constant reporting probability model (orange), which is significantly more trusting of $C_{+t}=0$.} 
\end{figure*}

The branching process perspective introduced in Eq. \ref{eq:bp} offers a useful platform to structure our thinking. Transmission trees grow and decay in the epidemiology characterized by $\mu_{+t}$ and $k_{+t}$. When it comes to introductions, we're concerned with trees rooted at some introduction time and then growing undetected before an observed transition from zero to nonzero cases. On the other hand, for elimination, we're interested in tree survival, that is the accumulating probability of spontaneous tree extinction in the wake of a transition from nonzero to zero cases.

In either situation, the key underlying idea is that we can grow transmission trees rooted at particular times, see how their size evolves, and then based on $r_{+t}$ we can calculate detection probabilities as a function of size and time. In the introduction context, we use these probabilities to estimate likely, unobserved root times, while in the elimination context, we use them to estimate unobserved extinction times. We should expect both of these times to be constrained since trees grow at the onset of outbreaks when $\mu_{+t}>1$, making longer periods of cryptic transmission less and less likely, and then on the tail end of outbreaks, trees spontaneously eliminate since $\mu_{+t}<1$.

Thankfully we don't actually have to sample trees and simulate this process. We leverage the formalism described in Ref. \citenum{thakkar2023generating} which is based on the classic branching process literature\cite{flajolet2009analytic}. At a high-level, we can define a probability generating function family, indexed by tree root times, which can be evaluated recursively since any tree is a root connected to smaller trees. That collection of generating functions lets us calculate tree statistics and survival probabilities exactly. The details are in Appendix \ref{app:D}.

In practice this means we can solve $\mathcal{P}$, specify Eqs. \ref{eq:trans_mod} and \ref{eq:bp}, and then precompute tree statistics as a function of root times across the whole series. With those inferences in hand, we walk along $C_{+t}$, highlight potential reintroduction and elimination transitions, and compose an overall probability of transmission interruption. 

The results of this process are visualized for London, Huddersfield, and Teignmouth in Fig. \ref{fig:elim}. For London (top panel), $C_{+t}$ is never zero, so the transmission interruption probability estimate is trivial. But as we move to smaller $+$communities, interruption probabilities grow and expose connections to the underlying transmission dynamics. Looking at Teignmouth specifically, probabilities are visibly asymmetric, tilted to the left, since transmission trees grow nonlinearly. We also see longer term trends, with higher peak probabilities in the 1950s. In Teignmouth, that was a period of lower birthrates and slower susceptibility accumulation, and reflecting that context, sustained periods of $C_{+t} = 0$ are longer in that decade.

To understand the role $r_{+t}$ plays, we can recompute the interruption probabilities in the constant reporting probability model (Appendix \ref{app:C}) with the same transmission tree properties. This comparison is visualized for Teignmouth in orange, and we find that if reporting rates were fixed at a constant average value, we'd be much more confident in measles spontaneously eliminating all the time, nearly independent of the underlying epidemiological context. 

The contrast in Teignmouth is striking, and it highlights that variation in $r_{+t}$ qualitatively changes the balance between cryptic transmission and transient elimination. Evidently, if we believe surveillance systems have lower sensitivity for sporadic importations and short transmission chains, we should expect the system to be at it's worst when it comes to verifying transmission interruption. The comparison at least encourages cautious thinking, and it emphasizes the need to better understand surveillance systems.



\section{\label{sec:conclusion}Conclusion}

Part of what makes zeros in cases interesting is that they force us to be clear about the interplay between local transmission, importation, and surveillance processes. In many situations, and often for good reason, researchers approach that issue with a mechanistic mindset, imposing structure on all three elements that they hope strikes the right balance between being interpretable, flexible, and tractable. 

One consequence of that thinking, since local transmission has a biological foundation and since population movement is potentially measurable, especially these days, is that we often encounter models with a lot of intricacy in the transmission and importation components. The surveillance process then has to be relatively rigid to compensate. 

The alternative line of thinking we've tried to articulate in this paper is that simplification of the interplay at a larger population scale effects the possibilities at finer scales. That observation meaningfully changes what's tractable, and it relaxes our need for structure. As a result, in the historical example we considered, we uncovered a set of estimates that highlighted interpretable dynamics in the surveillance system, which could be an important, neglected factor more generally and in modern contexts. It seems to warrant further study.

But as a concluding thought, it's worth noting that the two lines of reasoning aren't actually opposed. Depending on the details, if we have good measures of any of the components, they could be included in \ref{eq:nlp} or as lens to interpret the estimates, a more intentionally structured alternative to Eqs. \ref{eq:trans_mod} and \ref{eq:bp}. It's not all or nothing, and if we've learned anything, it's that we might benefit from thinking across scales more often.

\begin{acknowledgments}
This work grew a lot in discussion with my colleagues at the Gates Foundation. I want to thank Prashanth Selvaraj specifically for his thoughtful reading and comments.
\end{acknowledgments}

\appendix
\section{\label{app:A}An interior point method for regression with endemic constraints}

As described in the main text, $r_{-t} \geq \ell_{-t}$ is a simpler bound than $h_{t}(\mathbf{r}_{-t},S_{-0}) \geq 0$, since the former is linear in $r_{-t}$ and depends only on quantities at time $t$. If we want to solve \ref{eq:nlp} efficiently, we'll need to better understand how $h_{t}(\mathbf{r}_{-t},S_{-0})$ constrains $r_{-t}$. 

To even evaluate $h_{t}(\mathbf{r}_{-t},S_{-0})$ at some specific $\mathbf{r}_{-t}$, note first that $C_{-t} \sim \text{Binom}\{I_{-t},r_{-t}\}$ implies
\begin{align}
    \text{E}[I_{-t}|r_{-t},C_{-t}] &= \frac{C_{-t}+1}{r_{-t}} - 1,\label{eq:exp_I}\\
    \text{V}[I_{-t}|r_{-t},C_{-t}] &= \frac{1 - r_{-t}}{r_{-t}^2}\left(C_{-t}+1\right),\nonumber
\end{align}
so that $\sqrt{\text{V}[I_{-t}|r_{-t},C_{-t}]}/\text{E}[I_{-t}|r_{-t},C_{-t}] \sim \mathcal{O}\left(1/\sqrt{C_{-t}}\right)$. This means that $I_{-t}$ is sharp around $\text{E}[I_{-t}|r_{-t},C_{-t}]$ when the $-$community is large, and $\text{E}[S_{-t}I_{-t}|r_{-t},C_{-t}] \approx \text{E}[S_{-t}|r_{-t},C_{-t}]\text{E}[I_{-t}|r_{-t},C_{-t}]$ is a safe approximation. 

With susceptible dynamics like Eq. \ref{eq:S} in both the total and the $-$community, we then have
\begin{align}
    \text{E}[S_{-t}|\mathbf{r}_{-t},\mathbf{C}_{-t},\mathbf{B}_{-t}] = S_{-0} +\sum_{i=0}^{t-1} B_{-i} - \sum_{i=1}^{t}\text{E}[I_{-i}|r_{-i},C_{-i}],\label{eq:exp_S}
\end{align}
and the obvious analog for $\text{E}[S_{t}|\mathbf{r}_{t},\mathbf{C}_{t},\mathbf{B}_{t}]$. In practice, this means that given $\mathbf{r}_{-t}$ and initial condition $S_{-0}$, we can evaluate the expected $-$community states and compute differences from the full model to specify the expected $+$community states. This is why we concentrate on $\mathbf{r}_{-t}$ in \ref{eq:nlp}.

Working towards an iterative algorithm, consider $\mathbf{r}_{-t}^{(k+1)} =\mathbf{r}_{-t}^{(k)} + \mathbf{\Delta}_t$, where $\mathbf{\Delta}_t$ is a perturbation we seek to bound and then estimate. Inspired by the logic discussed below Eq. \ref{eq:ub}, we consider a situation where 
\begin{align*}
\mathbf{r}_{-t}^{(k+1)} = \mathbf{a}_{-t}^* \equiv (r_{-0}^{(k)}+\Delta_{0}, r_{-1}^{(k)}+\Delta_{1},...,r_{-t'-1}^{(k)}+\Delta_{t'-1},\ell_{-t'},...),
\end{align*}
with $t'$ marking the jump to the lower bound, after which $r_{-t}^{(k+1)}$ is arbitrary from the perspective of restricting $\Delta_{t\leq t'}$. For small $\Delta_{t}$, we take
\begin{align*}
h_{t'}(\mathbf{a}_{-t}^{*},S_{-0}) \approx h_{t'}((r_{-0}^{(k)}, r_{-1}^{(k)},...,r_{-t'-1}^{(k)},\ell_{-t'},...),S_{-0}) + \sum_{t}\left.\frac{\partial h_{t'}}{\partial r_{-t}}\right\rvert_{r_{-t}^{(k)}}\,\Delta_{t},
\end{align*}
making use of 
\begin{align*}
    \frac{\partial h_{t'}}{\partial r_{-t}} =
    \begin{cases}
        -\text{E}[I_{-t'}|\ell_{-t'},C_{-t'}]\left(\frac{\text{E}[I_{-t}|r_{-t},C_{-t}]+1}{r_{-t}}\right) & \text{if } 1 \leq t < t'\\
        0 & \text{otherwise}
    \end{cases}
\end{align*}
based on Eqs. \ref{eq:exp_I} and \ref{eq:exp_S}. Notice that $\partial h_{t'}/\partial r_{-t}$ is negative when it's nonzero, which means $h_{t'}(\mathbf{a}_{-t}^*,S_{-0})\geq 0$ is a collection of upper bounds on $\Delta_t$ indexed by $t'$. To find the strictest one, and so satisfy all of them, assume $\Delta_t = \Delta^*$ $\forall t$, with $\Delta^*$ determined by solving $\min_{t'}h_{t'}(\mathbf{a}_{-t}^{*},S_{-0}) = 0$. Thus, at a given iteration, we have the approximate constrained program,
\begin{align*}\tag{$\mathcal{P}^{(k)}$}\label{eq:nlp_k}
    \begin{split}
    \max_{\mathbf{\Delta_t}}& \quad \text{pr}(\mathbf{r}_{-t}^{(k)}+\mathbf{\Delta}_{t}\,|\,S_{-0},\mathbf{C}_{\pm t},\mathbf{B}_{\pm t})\\
    \text{subject to}&\quad \ell_{-t} \leq r_{-t}^{(k)} + \Delta_{t} \leq r_{-t}^{(k)} + \Delta^*\,\,\forall t,
    \end{split}
\end{align*}
which has only local, linear constraints, so-called box constraints. We can solve \ref{eq:nlp_k} with standard, gradient-based methods designed for these types of problems\cite{byrd1995limited}.

Summarizing an algorithm then, for fixed $S_{-0}$, start at $r_{-t}^{(0)} = \ell_{-t}$, which is feasible but suboptimal. Then compute $\Delta^*$ as line search step, solving $\min_{t'}h_{t'}(\mathbf{a}_{-t}^{*},S_{-0}) = 0$ with Newton's method, leveraging the linearization above. Next, solve \ref{eq:nlp_k} with L-BFGS, implemented in \texttt{scipy}. That defines $r_{-t}^{(1)}$, and the process repeats. In practice, we nest this interior-point method within a one-dimensional simplex search for the optimal $S_{-0}$. Throughout the paper, the method is implemented with a $10^{-10}$ tolerance for changes in the objective function. 

\section{\label{app:B}Autoregressive posterior probabilities}

The algorithm described above applies to any $\text{pr}(\mathbf{r}_{-t}\,|\,S_{-0},\mathbf{C}_{\pm t},\mathbf{B}_{\pm t})$, as long as it's differentiable for feasible $\mathbf{r}_{-t}$. We can structure possible choices by considering
\begin{align*}
    \text{pr}(\mathbf{r}_{-t}\,|\,S_{-0},\mathbf{C}_{\pm t},\mathbf{B}_{\pm t}) &\propto \text{pr}(\mathbf{C}_{\pm t}\,|\,S_{-0},\mathbf{r}_{-t},\mathbf{B}_{\pm t})\text{pr}(\mathbf{r}_{-t}\,|\,S_{-0},\mathbf{B}_{\pm t})\\
    &= \text{pr}(\mathbf{r}_{-t}\,|\,S_{-0},\mathbf{B}_{\pm t}) \int d\mathbf{I}_{\pm t} \,d\mathbf{S}_{\pm t}\, \text{pr}(\mathbf{C}_{\pm t},\mathbf{I}_{\pm t},\mathbf{S}_{\pm t}\,|\,S_{-0},\mathbf{r}_{-t},\mathbf{B}_{\pm t})
\end{align*}
where the first line is Bayes' theorem and the second introduces the infectious and susceptible populations in both communities. The joint distribution under the integrals can be hierarchically and sequentially decomposed to reflect meaningful conditional dependencies within a given model. That decomposition is sometimes represented graphically as a Bayesian network, and there are examples across the literature\cite{bishop2006pattern} on hidden Markov and other state-space models.

For us, our intention is to work towards the relationships in Eqs. \ref{eq:trans_minus} and \ref{eq:trans_plus}, given the susceptible and observation processes analogous to Eqs. \ref{eq:S} and \ref{eq:C}. As an example of the logic, consider
\begin{align*}
\text{pr}(\mathbf{C}_{\pm t},\mathbf{I}_{\pm t},\mathbf{S}_{\pm t}\,&|\,S_{-0},\mathbf{r}_{-t},\mathbf{B}_{\pm t})  = \text{pr}(\mathbf{C}_{\pm t},\mathbf{I}_{+ t},\mathbf{S}_{+ t}\,|\,\mathbf{I}_{- t},\mathbf{S}_{- t},S_{-0},\mathbf{r}_{-t},\mathbf{B}_{\pm t})\text{pr}(\mathbf{I}_{- t},\mathbf{S}_{- t}\,|\,S_{-0},\mathbf{r}_{-t},\mathbf{B}_{\pm t}),\\
&\approx \text{pr}(\mathbf{C}_{-t}\,|\,\mathbf{I}_{-t},\mathbf{r}_{-t}) \text{pr}(\mathbf{C}_{+t},\mathbf{I}_{+ t},\mathbf{S}_{+ t}\,|\,\mathbf{I}_{- t},\mathbf{S}_{- t},S_{-0},\mathbf{r}_{-t},\mathbf{B}_{\pm t})\text{pr}(\mathbf{I}_{- t},\mathbf{S}_{- t}\,|\,S_{-0},\mathbf{r}_{-t},\mathbf{B}_{\pm t}),
\end{align*}
where the first line is an exact choice and the second line is an assumption, based on the binomial observation model, that $\mathbf{C}_{-t}$ is independent of everything else given $\mathbf{I}_{-t}$ and $\mathbf{r}_{-t}$. We can make similar, sequential choices and assumptions. For example, for the final term above, we take
\begin{align*}
\text{pr}(\mathbf{I}_{- t},\mathbf{S}_{- t}\,|\,S_{-0},\mathbf{r}_{-t},\mathbf{B}_{\pm t}) &\approx \text{pr}(I_{-0})\prod_{t\geq0} \text{pr}(S_{-t+1}\,|\,S_{-0},I_{-0},...,I_{-t},\mathbf{B}_{- t})\text{pr}(I_{-t+1}\,|\,S_{-t},I_{-t}),
\end{align*}
which makes the Markov assumption in Eq. \ref{eq:trans_minus} and collects memory along the lines of Eq. \ref{eq:exp_S}. 

To work towards a tractable $\text{pr}(\mathbf{r}_{-t}\,|\,S_{-0},\mathbf{C}_{\pm t},\mathbf{B}_{\pm t})$, we string these types of assumptions together, similar to the treatment in Ref. \citenum{thakkar2022modeling}. Then, to take the integrals, approximate
\begin{align}
    \text{pr}(\mathbf{C}_{-t}\,|\,\mathbf{I}_{-t},\mathbf{r}_{-t}) &\approx \prod_{t} \delta\left(C_{-t} - \text{E}[C_{-t}|I_{-t},r_{-t}]\right),\label{eq:sharp_bin}
\end{align}
where $\delta(\cdot)$ is the Dirac-delta distribution, essentially assuming that binomial variance is negligible, and for our purposes here, $I_{-t} = C_{-t}/r_{-t}$ deterministically. Recalling also that everything above is implicitly conditional on the endemic model, so terms like $\text{pr}(I_{+t}|I_{-t},...) = \delta(I_t -(I_{+t} + I_{-t}))$, we ultimately take
\begin{align}
-\ln &\,\text{pr}(\mathbf{r}_{-t}\,|\,S_{-0},\mathbf{C}_{\pm t},\mathbf{B}_{\pm t}) \approx \text{const.} - \ln \text{pr}(\mathbf{r}_{-t}\,|\,S_{-0},\mathbf{B}_{\pm t})\\
&+ \frac{1}{2\text{V}[\tilde{\beta_t}]}\sum_{t}\left\{\left(\ln I_{-t+1} - \ln\frac{\beta_t}{F_-}S_{-t}I_{-t}\right)^2 + \mathbb{I}_{C_{+t}\neq0}\left(\ln I_{+t+1} - \ln\left[ \beta_tS_tI_t-\frac{\beta_t}{F_-}S_{-t}I_{-t}\right]\right)^2\right\}\label{eq:logpost}
\end{align}
where $\text{V}[\tilde{\beta}_t]$ comes from the endemic model and $\mathbb{I}_{C_{+t}\neq0}$ indicates times where the $+$community is reporting cases. All states above are evaluated under the approximation in Eq. \ref{eq:sharp_bin}. The prior $\text{pr}(\mathbf{r}_{-t}\,|\,S_{-0},\mathbf{B}_{\pm t})$ is the same as that from the endemic model as well, based on the age distribution of cases. Note however that in sensitivity testing, a constant prior negligibly effects the results. 

Eq. \ref{eq:logpost} defines the objective function used in \ref{eq:nlp} for all 60 cities, but hopefully it's clear that the derivation above could be adapted to other models or more refined approximations.

\section{\label{app:C}A rigid reporting model}
At a few places in the main text, we make comparisons to a constant reporting model. Specifically, we take
\begin{align*}
    I_{+t+1} = \gamma_tS_{+t}\left(I_{+t} + \theta\right),
\end{align*}
where $\gamma_t$ is a log-normal transmission rate and $\theta$ is a constant importation pressure. If we make (a not necessarily valid but common) assumption that $I_{+t}$ is sharp around $\text{E}[I_{+t}|r_{+t},C_{+t}]$, in analogy to the treatment of the $-$community in Appendix \ref{app:A}, we can take the log and rearrange to find
\begin{align}
    \ln\frac{C_{+t+1} + 1 - r_{+t}}{C_{+t} + 1 - r_{+t}} &= \ln\gamma_t\langle S_{+t}\rangle + \ln\left(1+\frac{S_{+t}-\langle S_{+t} \rangle}{\langle S_{+t} \rangle}\right) + \frac{r_{+t}\theta}{C_{+t} +1 - r_{+t}},\label{eq:smooth_r}
\end{align}
where we've assumed $r_{+t} \approx r_{+t+1}$, that is $r_{+t}$ varies slowly, and $\theta/I_{+t}$ is small. This relation is a starting point for a variety of smooth or covariate driven $r_{+t}$ models. In our case, we take $r_{+t} = r_{+}$ $\forall t$, and estimate both $r_{+}$ and the fluctuations $Z_{+t} = (S_{+t}-\langle S_{+t} \rangle)/\langle S_{+t} \rangle$ with the cumulative-cumulative regression method\cite{finkenstadt2000time,ferrari2008dynamics}, the same approach used to estimate the red dashed line in Fig. \ref{fig:estimates} for Huddersfield. In this constant $r_{+t}$ situation, Eq. \ref{eq:smooth_r} reduces to
\begin{align*}
    \ln\frac{C_{+t+1} + 1 - r_{+}}{C_{+t} + 1 - r_{+}} - \ln\left(1 + Z_{+t}\right)&= \ln x_t + \frac{r_{+}\theta}{C_{+t} +1 - r_{+}},
\end{align*}
where $x_t \equiv \gamma_t\langle S_{+t}\rangle$ is assumed to be annually periodic. This is a linear regression for $\ln x_t$ and $\theta$. After solving it, the seasonality profile is $\ln R_e = \ln x_t \lambda_t$, where $\lambda_t$ is the periodic time average of $1+ \theta/\text{E}[I_{+t}|r_{+t},C_{+t}]$. The corresponding volatility around $R_e$ can be approximated with standard linear regression and Taylor series methods.

\section{\label{app:D}Probabilities on transmission trees}
\begin{figure*}
\centering\includegraphics[width=\linewidth]{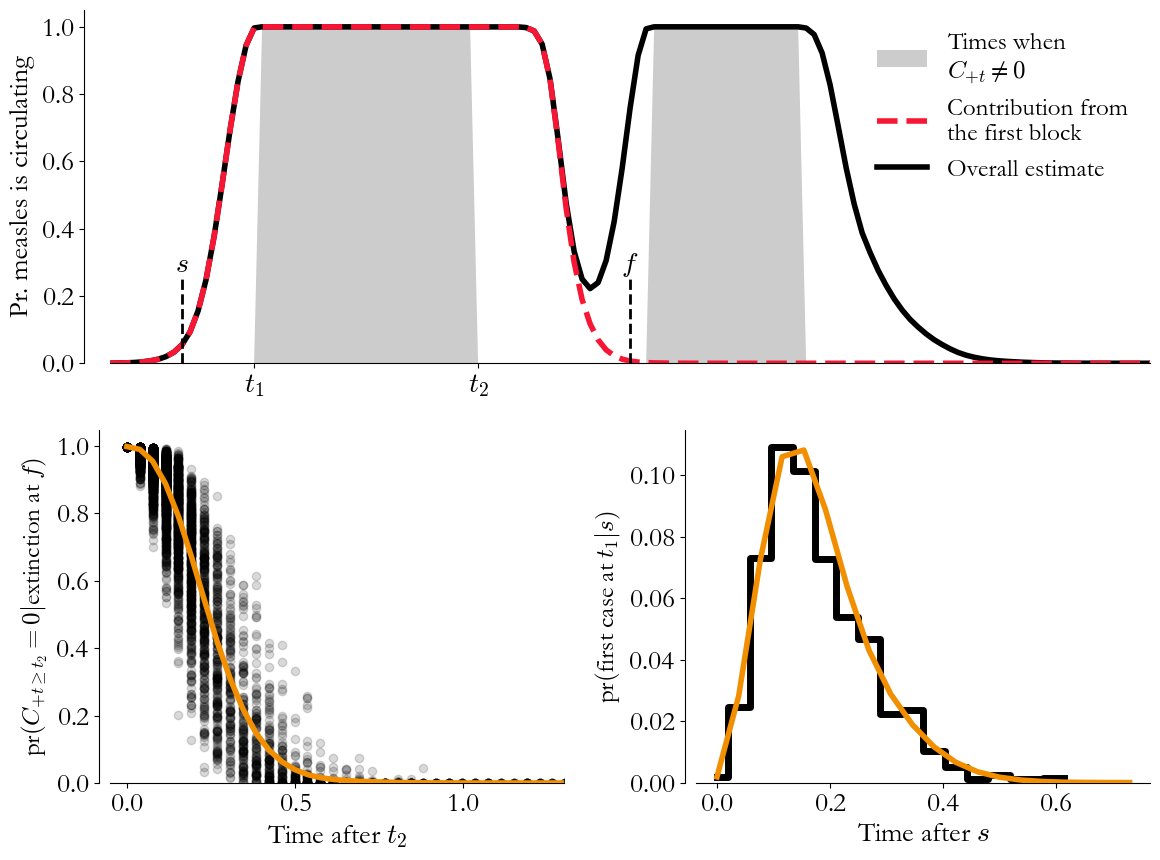}
\caption{\label{fig:forest} Signal processing with transmission trees. (top) Disjoint periods with $C_{+t}\neq0$ (grey) set up transitions to ($t_1$) and from ($t_2$) zero. The generating function equivalent to Eq. \ref{eq:bp} facilitates computing the probability of measles circulation from a single event pair (red dashed), which then we can compose over multiple blocks into an overall probability (black). (bottom) Sampling transmission trees and then sub-sampling nodes (black) to simulate reporting ($r_{+t} = r_{+} = 1/500$ here) validates the generating function-based calculations (orange).}
\end{figure*}

Consider the situation illustrated in Fig. \ref{fig:forest}, where $C_{+t} = 0$ up to time $t_1$, remains nonzero up to time $t_2$, and then transiently transitions back to $C_{+t} = 0$ just after $t_2$. Our strategy for estimating interruption probabilities will be to calculate the probability that transmission begins undetected at some start time $s \leq t_1$ and similarly that it continues undetected till some end time $f \geq t_2$. 

As mentioned in the main text, the branching process perspective introduced in Eq. \ref{eq:bp} helps us make progress. Following that literature\cite{flajolet2009analytic}, and along the lines we've developed for other, more complex use-cases\cite{thakkar2022covid,thakkar2023generating}, $T_{+t} \sim \text{NegBin}\{\mu_{+t},k_{+t}\}$ has probability generating function
\begin{align*}
    G_{T_{+t}}(z) &= \text{pr}(T_{+t}=0) + \text{pr}(T_{+t}=1)z + \text{pr}(T_{+t}=2)z^2 + ...\\
    &= \sum_{\ell \geq 0} \binom{k_{+t} + \ell - 1}{\ell} (p_{+t} z)^{\ell}(1-p_{+t})^{k_{+t}}\\
    &= \left(\frac{1-p_{+t}}{1-p_{+t}z}\right)^{k_{+t}}
\end{align*}
where $p_{+t} \equiv \mu_{+t}/(\mu_{+t}+k_{+t})$ and $z$ is an arbitrary complex number. Meanwhile, a generating function for transmission trees rooted at time $r$ can be represented formally as
\begin{align}
    T_r(\mathbf{z}) = \sum_{\tau \in \mathbb{T}_{r}} \text{pr}(\tau) z_{r}z_{r+1}^{n_{r+1}}\dotsm z_{\mathcal{T}}^{n_{\mathcal{T}}},\label{eq:ftgf}
\end{align}
where $\mathbf{z} = (z_{1},z_{2},...,z_{\mathcal{T}})$ is a complex vector across the length $\mathcal{T}$ time series. The sum above is over all possible trees $\tau$ in the set $\mathbb{T}_r$, with each tree characterized by its number of nodes, $n_t$, over time. Eq. \ref{eq:ftgf} is formal in the sense that we can't directly take the sum, but we do have
\begin{align}
    T_{r}(\mathbf{z}) = z_r G_{T_{+r+1}}(T_{r+1}(\mathbf{z})),\label{eq:rec}
\end{align}
a recursion relation capturing the idea that trees are rooted collections of smaller trees from the next infectious generation. This is a special case of the transmission forest theory\cite{thakkar2023generating}, with non-overlapping generations and with assumed isolated trees. Given the boundary condition $T_{\mathcal{T}}(\mathbf{z}) = z_\mathcal{T}$, Eq. \ref{eq:rec} offers a tractable, reverse-time path for evaluating Eq. \ref{eq:ftgf} at any $\mathbf{z}$ we choose.

To illustrate how the formalism above can be used, consider $\mathbf{z} = (1,1,...,1,0,0,...,0) \equiv \mathbf{\Theta}(t^*)$, a step function at time $t^*$ along $\mathbf{z}$. In that case, terms like $z_{r}z_{r+1}^{n_{r+1}}\dotsm z_{\mathcal{T}}^{n_{\mathcal{T}}}$ are only nonzero if $n_t = 0$ for $t>t^*$. Eq. \ref{eq:ftgf} implies that
\begin{align*}
    T_r(\mathbf{\Theta}(t^*)) &= \sum_{\tau \in \mathbb{T}_{r}} \text{pr}(\tau\,|\,n_{t>t^*} = 0)\\
    &= 1 - S_r(t^*),
\end{align*}
where $S_r(t^*)$ is the survival function for trees rooted at time $r$. Eq. \ref{eq:rec} then gives us an efficient recursive program to evaluate $T_r(\mathbf{\Theta}(t^*))$ across $r$ and $t^*$ and compute survival probabilities.

Along similar lines, consider $\mathbf{z} = (1-r_{+1}, 1 - r_{+2},...,1 - r_{+t^*},1,1,...,1) \equiv \mathbf{\Omega}(t^*)$, which would highlight trees that remain completely undetected up to $t^*$ and then are unconstrained. In the context of Fig. \ref{fig:forest}, that implies
\begin{align}
    \begin{split}
    \text{pr}(\text{first detection at }t_1 | s) &= \frac{T_{s}[\mathbf{\Omega}(t_1-1)] - T_{s}[\mathbf{\Omega}(t_1)]}{ S_{s}(t_1-1)},\\
    &= \frac{1}{S_{s}(t_1-1)}\sum_{\tau \in \mathbb{T}_{s}} \text{pr}(\tau)(1-r_{+s})(1-r_{+s+1})^{n_{s+1}}\dotsm[1 - (1-r_{+t_1})^{n_{t_1}}]\label{eq:intro}
    \end{split}
\end{align}
where the denominator conditions on survival after $t_1$. Since $T_r(\mathbf{\Omega}(t^*))$ is again recursively computable, Eq. \ref{eq:intro} can be evaluated across $s$ to find the most likely root time. This result is compared to simulated trees in the lower right of Fig. \ref{fig:forest}.

Meanwhile, on the tail end of an outbreak, we can think of $C_{+t_2}$ associated with an unknown number of roots, $m$. If $m=1$, then 
\begin{align*}
    \text{pr}(\text{extinction by }f|m=1,t_2)\equiv \varphi(f|t_2) = \frac{T_{t_2}[(1-\mathbf{r}_{+t})\otimes\mathbf{\Theta}(f)])}{T_{t_2}(1-\mathbf{r}_{+t})},
\end{align*}
where $\otimes$ is element-wise and the denominator conditions on $C_{+t>t_2}=0$. The numerator, $T_{t_2}[(1-\mathbf{r}_{+t})\otimes\mathbf{\Theta}(f)])$, is the probability that trees starting at $t_2$ go extinct before $f$, remaining completely undetected. It's change over $f$, conditional on survival till $f$, is compared to samples in Fig. \ref{fig:forest}.  

For general $m$, if the trees with distinct roots grow independently,
\begin{align*}
    \text{pr}(\text{extinction at }f|m,t_2) = \varphi(f|t_2)^m - \varphi(f-1|t_2)^m,
\end{align*}
since at least 1 tree needs to live till $f$. Marginalizing over $m$ with $C_{+t} \sim \text{Binom}\{I_{+t},r_{+t}\}$ gives
\begin{align}
    \begin{split}
    \text{pr}(\text{extinction at }f|t_2) &= \sum_{m} \text{pr}(I_{+t_2} = m |r_{+t_2},C_{+t_2}) \left[\varphi(f|t_2)^m - \varphi(f-1|t_2)^m\right]\\
    &= S(\varphi(f|t_2)) - S(\varphi(f-1|t_2))\label{eq:extin}
    \end{split}
\end{align}
where $S(x) = x^{C_{+t_2}}[r_{+t_2}/(1-x(1-r_{+t_2})]^{C_{+t_2}+1}$. Eq. \ref{eq:extin} allows us to calculate the most likely $f$ in periods after reporting ceases. 

Putting Eqs. \ref{eq:intro} and \ref{eq:extin} together, we can calculate the probability of ongoing transmission in this isolated example with a single transition to and from $C_{+t} \neq 0$. Specifically, before $t_1$, Bayes' theorem implies
\begin{align*}
    \text{pr}(s | \text{first detection at }t_1) \propto \text{pr}(\text{first detection at }t_1 | s) \text{pr}(s),
\end{align*}
which we can normalize directly if we take a constant prior over $s$. In advance of $t_1$, this let's us estimate the probability that $s$ has already passed. Meanwhile, after $t_2$, we can do the reverse for $f$, estimating the probability that $f$ hasn't passed yet. This idea is illustrated for a hypothetical example in red in Fig. \ref{fig:forest}, with $C_{+t_2}=6$. Notice that the estimate is asymmetric since we assume trees have $1$ root upon introduction, but we have to marginalize over possible roots for elimination, increasing our skepticism.

Real $C_{+t}$ have multiple sets of transitions. To estimate an overall probability of circulation, $\pi_t$, initialize
\begin{align*}
    \pi_t =
    \begin{cases}
        1 & \text{if } C_{+t} \neq 0\\
        0 & \text{if } C_{+t} = 0
    \end{cases}
\end{align*}
and group $C_{+t}$ into nonzero blocks, as illustrated in grey in Fig. \ref{fig:forest}. Moving along time, for each block, compute the corresponding circulation probability, $\psi_t$, associated with the transitions to and from zero. This is the red line in Fig. \ref{fig:forest} for the first block, as discussed above. Then, update
\begin{align*}
    \pi_t \leftarrow \pi_t + (1 - \pi_t)\psi_t,
\end{align*}
and move to the next block. Here the first term accounts for potential past cryptic circulation, and if that's not the case, the new evidence for circulation from the latest block. An example with 2 blocks is visualized in black in Fig. \ref{fig:forest}. Between periods of $C_{+t}\neq0$, evidence from the transitions overlaps, illustrating how this algorithm accumulates information over the full $C_{+t}$ series. 

\bibliography{references}

\end{document}